\documentclass[11pt]{article}
\usepackage{amsmath,amssymb,cite}
\makeatletter
\@addtoreset{equation}{section}
\makeatother

\def\ben{\begin{equation}}
\def\een{\end{equation}}

  \let\n=\nu  \let\p=\pi

\let\C=\Chi

\def\nn{\nonumber} \def\bd{\begin{document}} \def\ed{\end{document}}
\def\ds{\documentstyle} \let\fr=\frac \let\bl=\bigl \let\br=\bigr
\let\Br=\Bigr \let\Bl=\Bigl
\let\bm=\bibitem
\let\na=\nabla
\let\pa=\partial \let\ov=\overline
\newcommand{\be}{\begin{equation}}
\newcommand{\ee}{\end{equation}}
\def\ba{\begin{array}}
\def\ea{\end{array}}
\def\ft#1#2{{\textstyle{\frac{\scriptstyle #1}{\scriptstyle #2} } }}
\def\fft#1#2{{\frac{#1}{#2}}}
\def\del{\partial}
\def\vp{\varphi}
\def\sst#1{{\scriptscriptstyle #1}}
\def\oneone{\rlap 1\mkern4mu{\rm l}}
\def\td{\tilde}
\def\wtd{\widetilde}
\def\ie{{\it i.e.\ }}
\def\dalemb#1#2{{\vbox{\hrule height .#2pt
        \hbox{\vrule width.#2pt height#1pt \kern#1pt
                \vrule width.#2pt}
        \hrule height.#2pt}}}
\def\square{\mathord{\dalemb{6.8}{7}\hbox{\hskip1pt}}}
\newcommand{\ho}[1]{$\, ^{#1}$}
\newcommand{\hoch}[1]{$\, ^{#1}$}
\newcommand{\bea}{\begin{eqnarray}}
\newcommand{\eea}{\end{eqnarray}}
\newcommand{\ra}{\rightarrow}
\newcommand{\lra}{\longrightarrow}
\newcommand{\Lra}{\Leftrightarrow}
\newcommand{\bp}{\tilde \beta^\prime}
\newcommand{\tr}{{\rm tr} }
\newcommand{\Tr}{{\rm Tr} }
\def\0{{\sst{(0)}}}
\def\1{{\sst{(1)}}}
\def\2{{\sst{(2)}}}
\def\3{{\sst{(3)}}}
\def\4{{\sst{(4)}}}
\def\5{{\sst{(5)}}}
\def\6{{\sst{(6)}}}
\def\7{{\sst{(7)}}}
\def\8{{\sst{(8)}}}
\def\n{{\sst{(n)}}}
\def\cA{{{\cal A}}}
\def\cB{{{\cal B}}}
\def\cF{{{\cal F}}}
\def\cG{{{\cal G}}}
\def\cH{{{\cal H}}}
\def\tV{\widetilde V}
\def\tW{\widetilde W}
\def\tH{\widetilde H}
\def\tE{\widetilde E}
\def\tF{\widetilde F}
\def\tA{\widetilde A}
\def\im{{{\rm i}}}
\def\tY{{{\wtd Y}}}
\def\ep{{\epsilon}}
\def\vep{{\varepsilon}}
\def\bD{{{\bar D}}}
\def\R{{{\mathbb R}}}
\def\C{{{\mathbb C}}}
\def\H{{{\mathbb H}}}
\def\CP{{{\mathbb C}{\mathbb P}}}
\def\RP{{{\mathbb R}{\mathbb P}}}
\def\Z{{{\mathbb Z}}}
\def\bA{{{\mathbb A}}}
\def\bB{{{\mathbb B}}}
\def\bC{{{\mathbb C}}}
\def\bD{{{\mathbb D}}}
\def\bE{{{\mathbb E}}}
\def\bZ{{{\mathbb Z}}}
\def\Re{{{\frak{Re}}}}
\def\Im{{{\frak{Im}}}}
\def\cosec{{\,\hbox{cosec}\,}}
\def\Gm{{\Gamma_{\!\! -}}}
\def\Gp{{\Gamma_{\!\! +}}}
\def\stan{{standard }}
\def\nonstan{{supernumerary }}
\def\p{{\partial}}
\def\kdel#1{{\fft{\del}{\del#1}}}

\def\bog{{Bogomolny }}
\def\om{{\omega}}

\newcommand{\tamphys}{\it George and Cynthia Woods Mitchell  Institute
for Fundamental Physics and Astronomy,\\
Texas A\&M University, College Station, TX 77843, USA}

\newcommand{\auth}{
A. Bergman\hoch{\dagger}, H. L\"u\hoch{\dagger,\ddagger},
Jianwei Mei\hoch{\dagger} and C.N. Pope\hoch{\dagger,*}
}

\begin{document}

\begin{flushright}
\hfill{
MIFP-08-21\ \ \ USTC-ICTS-08-14}\\
%\hfill{
%\bf hep-th/yymmnnn}
\end{flushright}

\vspace{25pt}

\begin{center}

{\large {\bf AdS Wormholes}}

\vspace{25pt}
\auth

\vspace{10pt}
\hoch{\dagger}{\tamphys}

\vspace{10pt}

\hoch{\ddagger}{\it Interdisciplinary Center of Theoretical Studies,
USTC, Hefei, Anhui 230026, PRC}

\vspace{10pt}

\hoch{*}{\it  DAMTP, Centre for Mathematical Sciences,
 Cambridge University,\\  Wilberforce Road, Cambridge CB3 OWA, UK}

\vspace{40pt}

\underline{ABSTRACT}
\end{center}

We obtain a large class of smooth Lorentzian $p$-brane wormholes in
supergravities in various dimensions.  They connect two asymptotically
flat spacetimes.  In cases where there is no dilaton involved in the
solution, the wormhole can connect an AdS$_n\times S^m$ in one
asymptotic region to a flat spacetime in the other.  We obtain
explicit examples for $(n,m)=(4,7), (7,4), (5,5), (3,3), (3,2)$.
These geometries correspond to field theories with UV conformal fixed
points, and they undergo decompactification in the IR region.  In the
case of AdS$_3$, we compute the central charge of the
corresponding conformal field theory.

\vspace{15pt}

\thispagestyle{empty}

\pagebreak
%\voffset=0pt
%\setcounter{page}{1}
%
%\tableofcontents
%
%\addtocontents{toc}{\protect\setcounter{tocdepth}{2}}

%%%%%%%%%%%%%%%%%%%%%%%%%%%%%%%%%%%%%%%%

\section{Introduction}

Asymptotically AdS solutions in supergravities play an important
r\^ole in the AdS/CFT correspondence \cite{mald,wit,gkp}, since they
provide supergravity duals to quantum field theories with conformal
fixed points in the UV region.  In the bulk of such a solution, there
are limited possibilities.  There can be a black hole horizon with
non-zero (or zero) temperature, or there can be an AdS horizon of
different AdS radius, corresponding to a conformal field fixed point
in the IR region \cite{fgpw}.  A third possibility is that the
solution is solitonic, such as an R-charged AdS bubble solution in an
AdS gauged supergravity \cite{llm,clpbubble,llpv}.  Most likely, the
solution will have a naked singularity.  Examples include the large
class of AdS domain wall solutions with naked singularities
constructed in \cite{cglp,klt}, which are dual to the Coulomb branch
of the dual gauge theories.

A more intriguing situation is when there exists a wormhole in the
bulk that connects smoothly to different AdS boundaries.  In
Lorentzian signature such a geometry appears unlikely, and
disconnected boundaries can only be separated by horizons \cite{gsww}.
Thus the recent studies of wormholes in string theory and in the
context of the AdS/CFT correspondence have so far concentrated on
Euclidean-signature spaces \cite{mm,bcpvv,hop,bd,bcptv}.

In \cite{lumei}, Ricci-flat and charged Lorentzian wormholes in higher
dimensions were obtained.  These include the previously-known $D=5$
Ricci-flat case \cite{cd}.  The wormholes are smooth everywhere, and
connect two asympotically flat Minkowski spacetimes.  Although these
wormholes are not traversable geodesically (see \cite{aagc,aagccf} and 
\cite{lumei}),
it was demonstrated in \cite{lumei} that there exist traversable
accelerated timelike trajectories across the wormholes.

A class of magnetically-charged wormholes in $D=5$ supergravity was
also obtained in \cite{lumei}.  It was shown that for appropriate
choices of the parameters, the wormhole can connect an AdS$_3\times
S^2$ in one asympotic region to a Minkowski spacetime in the other.
This geometry then provides a supergravity dual of a two-dimensional
field theory at the boundary of the AdS$_3$.

In this paper, we begin in section 2 with a review of the Ricci-flat
wormhole solutions that were obtained in \cite{lumei}.  We then
construct $p$-brane wormhole solutions in section 3, supported by a
dilaton and $n$-form field strength.  In non-dilatonic cases, these
$p$-brane wormholes connect an AdS$_n\times S^m$ in one asymptotic
region to a flat spacetime in the other.  We obtain explicit examples
for $(n,m)=(4,7), (7,4), (5,5), (3,3), (3,2)$.  These geometries
correspond to field theories with UV conformal fixed points, which
undergo decompactification in the IR region.

In section 4, we study the AdS$_3$ wormhole obtained in \cite{lumei}
in detail and compute the central charge of the corresponding dual conformal
field theory.

In sections 5 and 6, we examine AdS$_5$,
AdS$_4$, AdS$_7$ and another AdS$_3$ wormhole in detail.
Included in these discussions is a calculation of the mass and momentum of
the configurations, as measured from the asymptotically AdS region.  To
do this, we make use of a construction of conserved charges in asymptotically
AdS spacetimes, which we summarise in an appendix.

     We conclude the paper in section 7.

\section{Ricci-Flat Wormholes in $D\ge 5$ Dimensions.}
\label{rfwh}

In this section, we review the Ricci-flat wormhole solutions
in general dimensions obtained in \cite{lumei}; they are given by
%%%%%%
\be
ds^2_D = (r^2 + a^2) d\Omega_{D-3}^2 + \fft{r^2 dr^2}{(r^2+a^2)
\sin^2 u} + \cos v (-dt^2+ dz^2) + 2\sin v\, dt dz\,,
\label{rfgenmet}
\ee
%%%%
where $v$ and $u$ are functions of $r$, given by
%%%%
\be
v=\sqrt{\fft{D-3}{2D-8}}\, (\pi - 2 u)\,,
\ee
%%%%
and
%%%
\be
u = \arctan\sqrt{\left(1+\fft{r^2}{a^2}\right)^{D-4} -1}
\,.\label{arctanform}
\ee
%%%%%%

   One can also rewrite the relation between $u$ and $r$ in the simpler
form
%%%%
\be
\cos^2u =\left (1 + \fft{r^2}{a^2}\right)^{4-D}\,.\label{cosform}
\ee
%%%
Note that $u=0$ when $r=0$.  Neither (\ref{arctanform}) nor
(\ref{cosform}) satisfactorily exhibits the fact that as $r$ passes
through zero, the {\it sign} of $u$ should be correlated with the sign
of $r$.  Instead, we can make this explicit by expanding the
expression in (\ref{arctanform}) and writing
%%%%%
\be
u= \arctan\left[ \fft{r}{a}\, \sqrt{\sum_{n=0}^{D-5} 
  \binom{D-4}{n+1} \, \Big(\fft{r}{a}\Big)^{2n}}
 \,\,  \right]\,.\label{arctan2form}
\ee
%%%%%
Thus we see that as $r$ ranges from $-\infty$ to $+\infty$, $u$ ranges
from $-\ft12\pi$ to $+\ft12\pi$.  For specific values of $D$, there
are sometimes simpler expressions for the relation between $u$ and
$r$, as we shall see later.  

   It is sometimes useful to use $u$, rather than $r$, as the radial 
coordinate.  The solution is then given by
%%%%
\be
ds_D^2 = \fft{a^2}{(\cos u)^{\ft2{D-4}}} \Big (\fft{du^2}{(D-4)^2 \cos^2 u}
+ d\Omega_{D-3}^2\Big) + \cos v\, (-dt^2 + dz^2) +
2\sin v\, dt\, dz\,,\label{rfgen2}
\ee
%%%%%
As was discussed in \cite{lumei}, the metric describes a smooth
wormhole in $D\ge 5$ dimensions that connects two flat asymptotic
spacetimes at $r\rightarrow \pm \infty$.  Note that the general
Ricci-flat wormhole metric (\ref{rfgenmet}) is related to that in
\cite{lumei} by a coordinate rotation in the $(t,z)$ plane.

      There are two asymptotic regions.  In the $r\rightarrow +\infty$
region, we have $u\rightarrow \ft12\pi$ and hence $v\rightarrow 0$.
It follows that the metric becomes
%%%%%
\be
ds^2=-dt^2 + dz^2 + dr^2 + r^2 d\Omega_{D-3}^2\,,
\ee
%%%%
which is a flat Minkowskian spacetime in $D$ dimensions.  In the
$r\rightarrow -\infty$ region, we ave $u\rightarrow -\ft12\pi$
and
%%%
\be
v\rightarrow 2\theta_0 \equiv 2\pi \sqrt{\fft{D-3}{2D-8}}\,.
\ee
%%%%
The metric becomes
%%%%%
\bea
ds^2&=& -\cos(2\theta_0) (dt^2-dz^2) + 2\sin (2\theta_0)\, dt dz +
dr^2 + r^2 d\Omega_{D-3}^2\,,\nn\\
&=& -d\td t^2 + d\td z^2 + dr^2 + r^2 d\Omega_{D-3}^2\,,
\eea
%%%%%
where
%%%%
\be
\begin{pmatrix}
\td t \cr \td z
\end{pmatrix} =
\begin{pmatrix}
\cos\theta_0 &
-\sin\theta_0 \cr \sin\theta_0 &
\cos\theta_0
\end{pmatrix}
\begin{pmatrix}
t\cr z
\end{pmatrix}\,.\label{asdef}
\ee
%%%
Thus the solutions connect smoothly two flat asymptotic spacetimes.
However, only for the case of $D=5$ do the two asymptotic regions have
the same time coordinate.  In higher dimensions, the notion of asymptotic
time is different in the two regions.

\section{General $p$-Brane Wormholes}
\label{pbranewh}

We construct charged wormholes as solutions to $\hat D$-dimensional Einstein
gravity coupled to an $n$-form field strength, together
with a dilaton.  The Lagrangian has the following general form
%%%%%
\be
{\cal L}_{\hat D}=\sqrt{-g} \Big(R- \ft12 (\del\phi)^2 -
\fft{1}{2\,n!} e^{\alpha\phi} F_\n^2\Big)\,.\label{genlag}
\ee
%%%%
where $F_\n=dA_{\sst{(n-1)}}$.  The constant $\alpha$ can be parameterised
as
%%%%
\be
\alpha^2= \Delta - \fft{2(n-1)(\hat D-n-1)}{\hat D-2}\,.\label{avalue}
\ee
%%%%%
The Langrangian (\ref{genlag}) is of the form that typically arises
as a truncation of the full Langrangian in many supergravities, 
with $\Delta$ being given by 
%%%%
\be
\Delta=\fft{4}{N}\,,
\ee
%%%%%
for integer $N$.  The values of $N$ that can arise depends on
the spacetime dimensions; they are classified in \cite{lpsoliton}.

   We may consider an electric ``$(n-2)$-brane wormhole'' in $\hat D$ spacetime
dimensions, based on the $D$-dimensional wormhole solution (\ref{rfgen2})
with
%%%%%
\be 
\hat D= D+n-3\,.
\ee
%%%%%
We therefore make as an ansatz for the
$\hat D$-dimensional metric and the dilaton field
%%%%%
\bea
ds_{\hat D}^2 &=& \fft{H^{\ft{(n-1)N}{(\hat D-2)}}}{(\cos u)^{\ft2{D-4}}}
\Big (\fft{a^2 du^2}{(D-4)^2 \cos^2 u} + a^2 d\Omega_{D-3}^2\Big)\nn\\
&& + H^{\ft{-(D-4)N}{(\hat D-2)}}\Big( \cos v\, (-dt^2 + dz^2) +
2\sin v\, dt\, dz + dx^i dx^i \Big)\,,\label{pbrane}\\
e^{\alpha\phi} &=& H^{2- \fft{(n-1)(D-4) N}{(\hat D-2)}}\,,\nn
\eea
%%%%%%
where the coordinates of the $D$-dimensional wormhole (\ref{rfgen2}) have
been augmented by $(n-3)$ additional world-volume coordinates $x^i$.
The function $H$ is assumed to depend only on the radial coordinate $u$.
For the field strength $F_\n$, we make the ansatz
%%%%%
\be
F_\n=
\sqrt{N}\,dt\wedge dz\wedge d^{n-3}x\wedge dH^{-1}\,,\label{Fsol}
\ee
%%%%%
Substituting into the equations of motion following from (\ref{genlag}), 
we find that they are all satisfied provided $H''=0$, and hence
$H$ is given by
%%%%%%
\be
H=c_0 - \fft{q}{a^{D-4}}\, u\,,
\ee
%%%%
where $c_0$ and $q$ are integration constants.  Without loss of
generality, let us take $q$ to be non-negative.  (Note that taking a
limit of $a\rightarrow 0$ leads to BPS p-brane solutions to the Lagrangian
(\ref{genlag}), obtained in general in \cite{stainless}.) It 
should again be emphasised that $r\leftrightarrow -r$ is not a symmetry, since
the expression for $u$ in terms of $r$ is defined by (\ref{arctan2form}),
showing that the signs of $u$ and $r$ are correlated.  The function
$H$ in general approaches a constant when $r\rightarrow \pm\infty$, given by
%%%%%
\bea
r\rightarrow +\infty:&& H\sim c_0 -\fft{\pi\,q}{2a^{D-4}} +
\fft{q}{r^{D-4}} + \cdots \,,\nn\\
r\rightarrow -\infty:&& H\sim c_0 + \fft{\pi\,q}{2a^{D-4}} -
(-1)^{[D/2]} \fft{q}{r^{D-4}}\,.
\eea
%%%%
Thus provided that $c_0> \pi\,q/(2a^{D-4})$, the $p$-brane wormholes
link two asymptotically flat spacetimes.  When $c_0=\ft12 \pi\,q a^{4-D}$,
for the non-dilatonic case $\alpha=0$, AdS wormholes can arise that link
AdS$\times$Sphere in the $r\rightarrow +\infty$ asymptotic region
to flat spacetime in the $r\rightarrow -\infty$ region.  We shall
discuss these solutions case by case in the following sections.

   Note that we can also consider ``magnetic $p$-brane wormholes,'' which
are equivalent to the previously-discussed electric cases, but constructed
using the $(\hat D-n)$-form dual of the $n$-form field strength $F_\n$. 
In other words, we can introduce the dual field strength
%%%%%
\be
\wtd F_{\sst{(\td n)}} = e^{\alpha\phi}\, {*F_\n}\,,
\ee
%%%%%
where $\td n=\hat D-n$, 
in terms of which the Lagrangian (\ref{genlag}) can be rewritten as
%%%%%
\be
{\cal L}_{\hat D}=\sqrt{-g} \Big(R- \ft12 (\del\phi)^2 -
\fft{1}{2\,\td n!} e^{-\alpha\phi} \wtd F_{\sst{(\td n)}}^2\Big)
\,.\label{genlagdual}
\ee
%%%%
where $\wtd F_{\sst{(\td n)}}=d\wtd A_{\sst{(\td n-1)}}$.  The
electric solution (\ref{pbrane}) of (\ref{genlag}), with $F_\n$ 
given by (\ref{Fsol}), can then 
then be reinterpreted as a magnetic solution of (\ref{genlagdual}), with 
$\wtd F_{\sst{(\td n)}}$ given by
%%%%%
\be
\wtd F_{\sst{(\td n)}} = \sqrt{N}\,(D-4)\, q\, \Omega_{D-3}\,.
\ee
%%%%%

\section{Magnetic String and AdS$_3$ Wormholes}
\label{ads3whsec}

In this section, we consider the magnetically-charged wormhole
solution in five-dimensional $U(1)^3$ supergravity that was obtained
in \cite{lumei}.  It was observed that for appropriate choice of the
parameters, the solution smoothly connect AdS$_3\times S^2$ in one
asymptotic region to the flat spacetime in the other.

We begin by reviewing the solution.  For simplicity, let us consider
the special case where all the three charges are equal.  The
corresponding minimal supergravity solution is given by
%%%%%%
\bea
ds_5^2&=&-H^{-1}
\Big(\fft{r^2-2as\, r -
a^2}{r^2 + a^2} dt^2 +\fft{4ac\,r}{r^2+a^2} dt\,dz -
\fft{r^2+2a s\, r-a^2}{r^2+a^2} dz^2\Big)\nn\\
&& + H^2 (dr^2 + (r^2+a^2)d\Omega_2^2)\,,\nn\\
F_\2&=&\sqrt3\,q\,\Omega_\2\,,\qquad H= c_0 - \fft{q}{a}
\arctan(\fft{r}{a})\,.
\label{d5sol1}
\eea
%%%%
It is straightforward to verify that this solution is contained in the
general form of $p$-brane wormhole (\ref{pbrane}) with $\hat D=D=5$,
$n=3$ and $N=3$.  Compared with (\ref{pbrane}), a boost
parameter $s=\sinh\beta$ ($c=\cosh\beta$) is also introduced, as in
 \cite{lumei}.  The solution describes a smooth charged
wormhole as long as
%%%
\be
c_0 \ge \fft{\pi q}{2a}\,.
\ee
%%%%
The coordinate $r$ runs from $-\infty$ to $+\infty$, corresponding to two
flat spacetimes when the above inequality holds.  Interesting things
happen when $ c_0=\pi q/(2a)$.  In this case, we have
%%%%%
\bea
r\rightarrow +\infty:&& H\sim \fft{q}{r} - \fft{q\,a^2}{3r^3} +
\fft{q\,a^4}{5r^5} - \fft{q\,a^6}{7r^7} + \cdots\,,\nn\\
r\rightarrow -\infty:&& H\sim \fft{q}{a}\Big( \pi +
\fft{a}{r} - \fft{a^3}{3r^3} + \fft{a^5}{5r^5} -
\fft{a^7}{7r^7} + \cdots \Big)\,.
\eea
%%%%%
Thus asymptotically as $r\rightarrow -\infty$, the spacetime is flat, whilst
as $r\rightarrow +\infty$, the spacetime is a direct
product of AdS$_3\times S^2$.

Since the size of the $S^2$ never vanishes, we can reduce the
solution on the $S^2$ and obtain a smooth solution in $D=3$.  Such
a breathing mode reduction was obtained in general dimensions
in \cite{instpaper}.  The reduction ansatz is given by
%%%%
\be
ds_5^2 = 
e^{2\alpha\varphi} ds_3^2 + e^{-\alpha \varphi} q^2 d\Omega_2^2\,,\qquad
F_\2 = \sqrt3\,q\,\Omega_\2\,,
\ee
%%%%
with $\alpha=1/\sqrt3$.  The $D=3$ system contains the metric and a
scalar, with the Lagrangian given by
%%%%
\be
{\cal L}_3 =\sqrt{-g} (R - \ft12 (\del\phi)^2 - V)\,,\qquad
V=-\fft{1}{2q^2} (4 e^{3\alpha \varphi} - 3 e^{4\alpha \varphi})\,.
\ee
%%%%%
The scalar potential contains an AdS fixed point $\varphi=0$.  The
resulting three-dimensional solution in the Einstein frame is given by
%%%%%
\bea
ds^2_3&=&q^{-4}\Big[(r^2 + a^2)^2 H^6 dr^2\label{ads3wh}\\
&&  - (r^2 + a^2) H^3
\Big((r^2-2as\, r -
a^2)dt^2 +4ac\,r\, dt\,dz -
(r^2+2a s\, r-a^2) dz^2\Big)\Big]\,,\nn\\
e^{-\alpha\phi} &=& q^{-2} (r^2 + a^2) H^2\,.\nn
\eea
%%%%%
In the asymptotic region $r\rightarrow -\infty$, we have that
$H$ is constant, and the solutions becomes 
%%%%%
\be
ds^2_3\sim r^4 (dr^2 - dt^2 + dz^2)\,,\qquad
\varphi\rightarrow -\infty\,.\label{afmet}
\ee
%%%%
The metric is locally flat\footnote{If we let $x^\mu=(t,z)$ 
and define $y^\mu= r^2\, x^\mu$ and $w= r^3/3$, the metric (\ref{afmet})
becomes
%%%%%
\bea
ds_3^2 = dw^2 + dy^\mu\, dy_\mu - \fft{4}{3w}\, y_\mu dy^\mu \, dw +
 \fft{4}{9w^2}\, y^\mu y_\mu\, dw^2\,,\nn
\eea
%%%%%
which is asymptotically flat when $|w|>>|y|$.}
for $r\rightarrow -\infty$; however, the
scalar describing the breathing mode of the internal $S^2$ diverges in
this limit. This breathing mode singularity is just a
lower-dimensional artifact, reflecting the fact that the radius of the
$S^2$ becomes infinite.  The system should really be lifted to five
dimensions in this limit.

%  We would like to compare the asymptotic behaviour
%of our solution to the BTZ black hole \cite{btz}, which is given by
%%%%%%
%\bea
%ds^2_3 &=& \fft{\ell^2}{r^2} dr^2 - \Big(1 - \fft{4M\ell^2}{r^2}
% + \fft{4\ell^2(M^2\ell^2 - J^2)}{r^4}\Big) \fft{r^2}{\ell^2}
%dt^2\nn\\
%&& + \Big(1 + \fft{4M\ell^2}{r^2} +
%\fft{4\ell^2 (M^2\ell^2-J^2)}{r^4}\Big) r^2 d\phi^2 -
%8 J\, dt\,d\phi\,,\label{btz}
%\eea
%%%%%
%where $M$ and $J$ are the mass and angular momentum of the black
%hole, and $\ell$ is the AdS$_3$ length.  (We are using the conventions of
%\cite{strom}, with Newton's constant $G=1$, here.)

%To make a comparison, We shall let $z$ be a compact circular
%coordinate, and make the following coordinate transformation (or
%renaming)
%%%%%%

    In the $r\rightarrow +\infty$ limit, the metric approaches
AdS$_3$. We would like to compute the central charge as in the work of
Brown and Henneaux \cite{BH}. As a first step, we make the following change of coordinates
\be
r\rightarrow \fft{r^2}{2q}\,,\qquad t\rightarrow \fft{t}{\sqrt{2}}\,,
\qquad z\rightarrow \fft{z}{\sqrt2}\,.
\ee
Including the subleading term at asymptotic
$+\infty$, the metric (\ref{ads3wh}) now has the following form:
%%%
\be
ds^2_3 = \fft{\ell^2}{r^2} dr^2 -
(1 - \fft{2a\,\ell\, s}{r^2}) \fft{r^2}{\ell^2} dt^2 +
(1 + \fft{2a\,\ell\,s}{r^2}) \fft{r^2}{\ell^2} dz^2 - \fft{4a c}{\ell}\,dt\, dz\,,
\ee
%%%%%
where $\ell=2q$.  If we had considered the general three unequal
charge solution, we would have $\ell=2(q_1q_2q_3)^{1/3}$. The traditional form
of the Poincar\'e patch of the AdS$_3$ metric is:
\begin{equation}
\label{poin}
ds^2_3 = \frac{R^2}{r^2}dr^2 + \frac{r^2}{R^2} \left(dz^2 - dt^2\right)\ .
\end{equation}

% Comparing
%this with the BTZ black hole, and following the discussion in
%\cite{bhtz,strom}, we have
%%%%%
%\be
%M=\fft{1}{\ell} (L_0 + \bar L_0) =\fft{a\,s}{2\ell}\,,\qquad
%J= L_0 - \bar L_0 = \fft{a\,c}{2}\,,
%\ee
%%%%%
%Thus we have a dual comformal
%field theory with
%%%%
%\be
%L_0=\ft14 (c+s)a \,,\qquad \bar L_0  = -
%\ft14 (c-s)a\,.
%\ee
%%%%
We can now directly compare with, say, \cite{BK, Henningson:1998gx} and see
that the central charge of the system is
%%%%
\be
C=\ft32 R= \ft32\ell\,.
\ee
%%%%%
Here, we use $C$ rather than the more conventional $c$ to denote the central 
charge, to avoid confusion with the short-hand notation $c=\cosh\beta$
of this paper. 

%To have negative $\bar L_0$
%suggests negative norm states.  Yet on the other hand, the dual
%supergravity solution is smooth.  It appears that the negative normal
%states may be responsible for the decompactification of the system in
%the IR region.

%The deviation of our wormhole solution from the form of the BTZ black
%hole starts at the $1/r^8$
%order.  The AdS$_3$ wormhole up to this order is given by
%%%%%
%\bea
%ds^2_3&=& \fft{\ell^2}{r^2} \Big(1 - \fft{32\alpha}{15 r^8}\Big)dr^2
%-\fft{r^2}{\ell^2} \Big(1 - \fft{4M\ell^2}{r^2} +
%\fft{4\ell^2 (M^2\ell^2-J^2)}{r^4} -
%\fft{16\alpha}{15r^8}\Big)
%dt^2\nn\\
%&&+r^2\Big(1 + \fft{4M\ell^2}{r^2} +
%\fft{4\ell^2(M^2\ell^2 - J^2)}{r^4} -\fft{16\alpha}{15r^8}
%\Big)d\phi^2\nn\\
%&&
%-8 J \Big(1 - \fft{16\alpha}{15r^8}\Big) dt\,d\phi +
%\cdots \,,
%\eea
%%%%%%%
%where $\alpha = \ell^4 (M^2\ell^2 - J^2)^2$.  There exists a BPS limit
%$M\ell=J$.  This can be achieved by sending $\beta$ to infinity and
%$a$ to zero, but keeping $a\,s$ to be finite and non-vanishing.  The
%resulting metric becomes the extremal BTZ black hole.

%It is necessary to identify $z$ as circular coordinate, so as to have
%asymptotic BTZ structure.  The consequence of this identification is
%that the solution has naked CTC's, and the wormhole is also a time
%machine.  However, the solution has no Killing horizon, and hence the
%asymptotic time coordinate $t$ does not have to be periodically
%identified, which is preferable to the previous time-machine solutions
%obtained in \cite{timemachine}.

\section{D3-brane and AdS$_5$ Wormhole}
\label{ads5wh}

It is perhaps more interesting to obtain an AdS$_5$ wormhole in type
IIB theory, which would be expected to be dual to certain
four-dimensional Yang-Mills theory.  Since AdS$_5$ appears naturally
in the type IIB theory in AdS$_5\times S^5$, the near horizon geometry
of the D3-brane, we consider the D3-brane wormhole solution.  From
(\ref{pbrane}) with $D=8$, $n=5$ and $N=1$, we find that the D3-brane
wormhole solution is given by
%%%%
\bea
ds^2_{10} &=& \Big(\fft{H}{\cos u}\Big)^{1/2}\, 
  \Big[ a^2 d\Omega_5^2 + \fft{a^2 du^2}{16 \cos^2 u} \Big]\nn\\
&& + H^{-1/2}\Big(\cos v\, (-dt^2+dz^2) + 2 \sin v
\, dt\,dz + dx_1^2 + dx_2^2\Big)\,,\nn\\
F_\5&=&G_\5 + {*_{10} G_\5}\,,\qquad
G_\5=dt\wedge dz\wedge dx_1\wedge dx_2\wedge dH^{-1}\,,\label{d3brane}\\
H&=&c_0 - \fft{q}{a^4}\, u\,,\quad
u=2\arcsin\Big( 
       \fft{r(r^2+2a^2)^{1/2}}{\sqrt2 (r^2+a^2)}\Big)\,,
\quad v=\sqrt{\ft58} (\pi-2u)\,.\nn
\eea
Choosing the integration constant $c_0=\pi\,q/(2a^4)$, we have
%%%%
\bea
r\rightarrow +\infty:&& H \sim \fft{q}{r^4} -\fft{2a^2q}{r^6} +
\cdots\,,\nn\\
r\rightarrow -\infty:&& H\sim \fft{\pi\,q}{a^4} - \fft{q}{r^4} +
\cdots\,.
\eea
%%%
Thus we have constructed a wormhole solution that is asymptically
AdS$_5\times S^5$ when $r\rightarrow +\infty$ and flat when
$r\rightarrow -\infty$.

   Since the solution obtained above is spherically symmetric, it can be
dimensionally reduced on $S^5$ to give a solution in five dimensions.
Using the results in \cite{instpaper}, we can reduce to the five-dimensional
metric $ds_5^2$ given by
%%%%%
\bea
ds_{10}^2 &=& e^{2\alpha\varphi}\, ds_5^2 + 
             e^{-\ft{6\alpha}{5} \varphi} \ell^2 d\Omega_5^2\,,\nn\\
G_\5&=& 4 \ell^4\, \Omega_\5\,,
\eea
%%%%%
where $\alpha= \sqrt{5/48}$ and $\ell\equiv q^{1/4}$ is the
AdS$_5$ length.  The five-dimensional
system is then Einstein gravity coupled to a dilaton with
a scalar potential, namely
%%%%
\be
{\cal L}_5 = \sqrt{-g} \Big(R-\ft12 (\del\varphi)^2 - V\Big)\,,
\ee
%%%
with the scalar potential given by
%%%%
\be
V=-\fft{4}{\ell^2} (5 e^{\ft{16}{5}\alpha\varphi} - 2
e^{8\alpha\varphi})\,.\label{d5pot}
\ee
%%%

     Performing the $S^5$ reduction on the solution, we find
%%%%%
\bea
ds^2_5 &=&\Big(\fft{a}{\ell}\Big)^{\ft{10}3}\Big\{
\Big(\fft{H}{\cos u}\Big)^{4/3}\, \fft{a^2 du^2}{16\cos^2 u}\nn\\
&& \quad\quad\qquad +
 \fft{H^{1/3}}{\cos^{5/6} u}\, 
  \Big(\cos v\, (-dt^2+dz^2) + 2 \sin v
\, dt\,dz + dx_1^2 + dx_2^2\Big)\Big\}\,.\nn\\
e^{-\ft{6\alpha}{5}\varphi} &=& \fft{a^2}{\ell^2} \sqrt{\fft{H}{\cos u}}\,.
\eea
%%%%%
There is no metric singularity as $u$ ranges from $-\ft12\pi$ to
$\ft12\pi$, corresponding to $r$ ranging from $-\infty$ to $+\infty$.
In the asymptotic region $r\rightarrow -\infty$, the solution becomes
%%%%
\bea
ds^2_5&\sim & \pi^{4/3} \Big(\fft{r^2}{\ell^2}\Big)^{\ft{5}3}\,
\Big(\fft{\ell}{a}\Big)^{16/3}
\Big[dr^2 + \fft1{\pi} \Big(\fft{a}{\ell}\Big)^4 (-d\td t^2 + d\td z^2 +
dx_1^2 + dx_2^2) \Big]\,,\nn\\
\varphi &\sim& -\infty\,,
\eea
%%%%%
where $\td t$ and $\td z$ are defined in (\ref{asdef}).  The metric is
locally flat in this limit, in the sense that the Riemann tensor tends to
zero as $r$ approaches $-\infty$ (see footnote 1).  
The dilaton is singular in the limit,
but the scalar potential (\ref{d5pot}) goes to zero.  Such a
singularity, in which the potential is bounded above, is called a ``good
singularity'' in \cite{gubser}.  However, the situation considered in
\cite{gubser} is when such a singularity occurs at a 
finite $r=r_0$, where $g_{tt}\rightarrow 0$, corresponding to
non-trivial infrared physics in the dual field theory.  In our case,
only the scalar becomes singular as
$r\rightarrow -\infty$, with $g_{tt}\sim \infty$.  From the
supergravity point of view, this is clearly a ``good singularity,'' and
is nothing but an artifact of the dimensional reduction.  It implies
that the system decompactifies into ten dimensions.

In the asymptotic region $r\rightarrow +\infty$, the solution becomes,
%%%%
\bea
ds_5^2 &\sim& \fft{\ell^2}{r^2} dr^2 + \fft{r^2}{\ell^2} 
(-dt^2 + dz^2 + dx_1^2 + dx_2^2)\,,\nn\\
\varphi &\sim& 0\,.
\eea
%%%
(Note that in the limit of $a\rightarrow 0$, the solution becomes
literally AdS$_5$ for all $r$.)  

    The scalar potential (\ref{d5pot}) 
tends to zero as $\varphi$ goes $-\infty$, and diverges to $+\infty$ as
$\varphi$ goes to $+\infty$.  There is a minimum, $V_{\rm min}=-12/\ell^2$,
which occurs at $\varphi=0$. The wormhole solutions corresponds to
$\varphi$ traversing from $\varphi=-\infty$ in the asymptotically locally
flat region to $\varphi=0$ in the asymptotically AdS region, with 
$V_{\rm min}$ determining the cosmological constant of AdS$_5$.  

      Including subleading terms, the metric in the asymptotically
AdS region is given by
%%%%
\bea
ds_5^2 &\sim& \fft{\ell^2}{r^2} (1 - \fft{2a^2}{r^2}) dr^2 +
\fft{r^2}{\ell^2} (1 + \fft{a^2}{r^2}) [-dt^2 + dz^2 +
dx_1^2 + dx_2^2 + \fft{\sqrt{10} a^4}{r^4} dt dz]\,,\nn\\
\varphi &\sim& -\fft{5\sqrt{\ft53} a^8}{r^8}\,.
\eea
%%%%%
If we express the metric near $r=+\infty$ 
in terms of a radial coordinate $\rho$, for which
the metric in the radial direction is $\ell^2 d\rho^2/\rho^2$, then
the solution looks like
%%%%%
\bea
ds^2_5 &=& \fft{\ell^2}{\rho^2} d\rho^2 + \fft{\rho^2}{\ell^2}\, 
\Big[ \Big(1-\fft{25 a^8}{24\rho^8}\Big) (-dt^2 + dz^2) +
\fft{\sqrt{10}\, a^4}{\rho^4}\, dt dz \nn\\
&&\qquad\qquad\qquad 
+ \Big(1+ \fft{5 a^8}{24\rho^8}\Big)
 (dx_1^2+dx_2^2)\Big] +\cdots
\eea
%%%%%
Using the energy and momentum formulae obtained in appendix  
\ref{mass}, we can straightforwardly to obtain the mass and linear
momentum per unit 3-volume spanned over $(z, x_1, x_2)$ for the
AdS$_5$ wormhole, given by
%%%%
\be
E=0\,,\qquad P=\fft{\sqrt{10}\, a^4}{8\pi \ell^5}\,.
\ee
%%%%
Of course, we can boost the system along the $(t, z)$ direction
and obtain a non-zero mass; however, we shall always have $E^2-P^2
  = -25 a^8/(16\pi^2 \ell^{10}) <0$.

\section{Further AdS Wormholes}
\label{morewh}

\subsection{M2-brane and AdS$_4$ wormhole}

   We can obtain an M2-brane wormhole solution of eleven-dimensional
supergravity, given by (\ref{pbrane}) with $D=10$, $n=4$, 
$N=1$:
%%%%
\bea
ds_{11}^2 &=& \Big(\fft{H}{\cos u}\Big)^{1/3}\, \Big[ a^2 d\Omega_7^2 
  + \fft{a^2\, du^2}{36\cos^2 u}\Big]\nn\\
&& +H^{-2/3}\, \Big(\cos v\,(-dt^2 + dz^2) +2 \sin v\, dt dz + dx^2\Big)\,,
\nn\\
A &=&  \fft{1}{H}dt\wedge dz\wedge dx\,,\qquad
H= c_0- \fft{q}{a^6} u\,,\qquad
v=\sqrt{\ft7{12}} (\pi-2u)\,.
\label{ads4sol}
\eea
%%%%%
The coordinate $u$ is related to the original $r$ coordinate by 
%%%%%
\be
u=\arctan\sqrt{\Big(1 + \fft{r^2}{a^2}\Big)^6-1}\,,
\ee
%%%%%
or, in other words,
%%%%%
\be
\cos u = \Big(1+\fft{r^2}{a^2}\Big)^{-3}\,,\qquad
\sin\ft12 u = \fft{r (r^4+3 r^2 a^2 + 3 a^4)^{1/2}}{
  \sqrt2 (r^2+a^2)^{3/2}}\,.
\ee
%%%%%
As $r$ ranges from $-\infty$ to $+\infty$, $u$ ranges from $-\ft12\pi$ 
to $+\ft12 \pi$.  

    Choosing $c_0=\pi q/(2 a^6)$, we find that 
$H$ has the asymptotic forms
%%%%%
\bea
r\rightarrow+\infty: && H=\fft{q}{r^6} -\fft{3a^2 q}{r^8} + \cdots\,,\nn\\
r\rightarrow -\infty:&& H= \fft{\pi q}{a^6} - \fft{q}{r^6}
      +  \fft{3a^2 q}{r^8} +\cdots\,.
\eea
%%%%%
The metric approaches AdS$_4\times S^7$ near $r=+\infty$, while it
becomes flat as $r$ approaches $-\infty$.

   A breathing-mode reduction of eleven-dimensional supergravity on $S^7$,
in which the metric and 4-form are written as
%%%%%
\bea
ds_{11}^2 &=& e^{2\alpha\varphi}\, ds_4^2 + 
 (2\ell)^2 \, e^{-\ft{4\alpha}{7}\varphi}\, d\Omega_7^2\,,\nn\\
{*F_\4} &=& 6 (2\ell)^6\, \Omega_7\,,
\eea
%%%%%%
where $\alpha=\sqrt7/6$, yields \cite{instpaper} the four-dimensional
bosonic Lagrangian
%%%%%
\be
{\cal L}_4 = \sqrt{-g}\, \Big( R -\ft12 (\del\varphi)^2 -V\Big)\,,
\qquad V= -\fft{3}{2\ell^2}\, \Big( 7 e^{\ft{18\alpha}{7}\varphi} -
                       3 e^{6\alpha\varphi}\Big)\,.
\ee
%%%%%
Reducing the solution (\ref{ads4sol}), with $2\ell=q^{1/6}$, therefore
gives the four-dimensional solution
%%%%%
\bea
ds_4^2 &=& \Big(\fft{a}{2\ell}\Big)^7\, \Big\{ \Big(\fft{H}{\cos u}\Big)^{3/2}
  \, \fft{a^2 du^2}{36 \cos^2 u} \nn\\
&&\qquad\qquad  +\fft{H^{1/2}}{\cos^{7/6} u} \Big(
  \cos v\, (-dt^2+dz^2) + 2 \sin v\, dt dz + dx^2\Big)\Big\}\,,\nn\\
e^{-\ft{4\alpha}{7}\varphi} &=& \fft{a^2}{4\ell^2}\, 
 \Big(\fft{H}{\cos u}\Big)^{1/3}\,.
\eea
%%%%%
In the limit of $r\rightarrow \infty$, the metric including
the subleading terms is given by
%%%%%
\be
ds_4^2 = \fft{\ell^2}{\td r^2} (1 -\fft{a^2}{2\ell\td r}) d\td r^2 +
\fft{\td r^2}{\ell^2} (1 + \fft{a^2}{2\ell\td r}) (-dt^2 + dz^2 +
dx^2 +  \fft{\sqrt{7/3}\, a^6}{\ell^3 \td r^3} dt dz) +\cdots\,,
\ee
%%%%
where $\td r = r^2/(4\ell)$.  Expressed instead in terms of a radial variable
$\rho$ for which the radial term is exactly $\ell^2 d\rho^2/\rho^2$, the 
expansion at large $r$ takes the form
%%%%%
\be
ds_4^2 = \fft{\ell^2 d\rho^2}{\rho^2} + \fft{\rho^2}{\ell^2}\,\Big[ 
  \Big(1-\fft{7 a^{12}}{2^{15}\ell^6 \rho^6}\Big) (-dt^2+ dz^2)
 +  \Big(1+\fft{7 a^{12}}{3\cdot 2^{15} \ell^6\rho^6}\Big) dx^2 +
  \fft{\sqrt{7/3}\, a^6}{32\ell^3 \rho^3}\, dt dz \Big]+\cdots\,.
\ee
%%%%%

   Using the formulae in appendix \ref{mass} 
for the energy and $z$-momentum per unit area in
the $(z,x)$ plane, we find
%%%%%
\be
E=0\,,\qquad P= \fft{\sqrt{21}\, a^6}{8\pi (2\ell)^7}\,.
\ee
%%%%%

\subsection{M5-brane and AdS$_7$ wormhole}

   There is also an M5-brane wormhole solution in eleven-dimensional
supergravity, given by taking $D=7$, $n=7$, $N=1$ in 
(\ref{pbrane}):
%%%%%
\bea
ds_{11}^2 &=& (\fft{H}{\cos u}\Big)^{2/3} \Big[a^2 d\Omega_4^2 +
    \fft{a^2 du^2}{9\cos^2 u}\Big] \nn\\
&&+ H^{-1/3} 
  \Big( \cos v (-dt^2 + dz^2) +
  2\sin v\, dt dz + dx^i dx^i\Big)\,,\nn\\
F_4 &=& 3 q \, \Omega_4\,,\qquad
H = c_0 -\fft{q}{a^3}\, u\,,\label{ads7sol}
\eea
%%%%%
where 
%%%%%
\be
v= \fft{2(\pi -2u)}{\sqrt6}\,.
\ee
%%%%%
The index $i$ ranges over $1\le i\le 4$.

   The coordinate 
$u$ is related to the original radial coordinate $r$ by
%%%%%
\be
\cos u= \Big(1+\fft{r^2}{a^2}\Big)^{-3/2}\,,\qquad
 \sin u= \fft{r(r^4+ 3r^2 a^2+ 3 a^4)^{1/2}}{(r^2+a^2)^{3/2}}\,.
\ee
%%%%%
As usual, $r$ ranges from $-\infty$ to $+\infty$, implying that $u$
ranges from $-\ft12\pi$ to $+\ft12\pi$.  

    With the choice  $c_0=\pi q/(2a^3)$,
the function $H$ tends to zero at $r=+\infty$ and it tends to
a constant at $r=-\infty$:
%%%%%
\bea
r\rightarrow +\infty:&& H = \fft{q}{r^3} - \fft{3 q a^2}{2 r^5}
+\cdots\,,\nn\\
r\rightarrow -\infty:&& H=\fft{\pi q}{a^3} - \fft{q}{r^3} + 
\fft{3 q a^2}{2 r^5} +\cdots\,.
\eea
%%%%%
The metric approaches AdS$_7\times S^4$ near $r=+\infty$, while it becomes
flat as $r$ approaches $-\infty$.

  The breathing-mode reduction of eleven-dimensional supergravity on
$S^4$, using the ansatz
%%%%%
\bea
ds_{11}^2 &=& e^{2\alpha\varphi}\, ds_7^2 + (\ft12\ell)^2\, 
e^{-\ft52\alpha \varphi}
\, d\Omega_4^2\,,\nn\\
F_4 &=& 3(\ft12\ell)^3\, \Omega_4\,,
\eea
%%%%%
with $\alpha=2/(3\sqrt{10})$, yields the seven-dimensional bosonic Lagrangian
%%%%
\be
{\cal L}_7= \sqrt{-g}\Big( R -\ft12 (\del\varphi)^2 -V\Big)\,,\qquad
V=- \fft{6}{\ell^2}\Big( 8 e^{\ft92\alpha\varphi} -
          3 e^{12\alpha\varphi}\Big)\,.
\ee
%%%%%
The solution (\ref{ads7sol}), with $\ft12\ell=q^{1/3}$, reduces to give
%%%%%
\bea
ds_7^2 &=& \Big(\fft{2a}{\ell}\Big)^{8/5}\, 
  \Big\{ \Big(\fft{H}{\cos u}\Big)^{6/5}
  \, \fft{a^2 du^2}{9 \cos^2 u} \nn\\
&&\qquad\qquad + \fft{H^{1/5}}{\cos^{8/15} u} \Big(
  \cos v\, (-dt^2+dz^2) + 2 \sin v\, dt dz + dx^i dx^i\Big)\Big\}\,,\nn\\
e^{-\ft{5\alpha}{2}\varphi} &=& \fft{4a^2}{\ell^2}\,
 \Big(\fft{H}{\cos u}\Big)^{2/3}\,.
\eea
%%%%%
Taking the limit $r\rightarrow
\infty$, the metric up to sub-leading order terms is
%%%%%
\be
ds_7^2 = \Big(1 - \fft{8a^2\ell^2}{\td r^4}\Big)
\fft{\ell^2 d\td r^2}{\td r^2} + \fft{\td r^2}{\ell^2}
  \Big(1 + \fft{2a^2\ell^2}{\td r^4}\Big)\Big(-dt^2 + dz^2 + dx^i dx^i +
\fft{32\sqrt{2/3}\,a^3 \ell^3}{\td r^6}dt dz\Big)+\cdots\,,
\ee
%%%%
where $\td r$ is given by $r=\td r^2/(2\ell)$.  Expressed in terms
of a radial coordinate $\rho$ for which the radial term in the metric
is exactly $\ell^2 d\rho^2/\rho^2$, the expansion takes the form
%%%%%
\bea
ds_7^2 &=& \fft{\ell^2 d\rho^2}{\rho^2} + 
 \fft{\rho^2}{\ell^2}\Big[
  \Big(1-\fft{384a^6 \ell^6}{5\rho^{12}}\Big) (-dt^2 +
dz^2) + \Big(1+\fft{128 a^6\ell^6}{15\rho^{12}}\Big)
 dx^i dx^i\nn\\ 
&&\qquad \qquad \qquad 
  +\fft{32\sqrt{2/3} \, a^3\ell^3}{\rho^6}\, dt dz\Big] +\cdots\,.
\eea
%%%%%

  The energy and $z$-momentum per unit 5-volume spanned by $(z,x^i)$,
calculated using the results in appendix \ref{mass}, are given by
%%%%%
\be
E=0\,,\qquad P= \fft{2\sqrt6\, a^3}{\pi\ell^4}\,.
\ee
%%%%%

\subsection{Dyonic string and AdS$_3$ wormholes}

The dyonic string is a six dimensional solution supported by
a 3-form field strength, corresponding to the Lagrangian
(\ref{genlag}), but with $\hat D=6$, $n=3$ and $N=1$.
The solution can be lifted to $D=10$ and viewed as either
the D1/D5 system or the NS-NS-1/NS-NS-5 system.

There are two dyonic string wormhole 
solutions.  The first one can be obtained
by lifting the magnetic string solution of the $U(1)^3$ theory
obtained in \cite{lumei} to six dimensions.  It is given by
%%%%%
\bea
ds^2_6 &=&(H_e H_m)^{-\ft12} \Big(\fft{r^2-2as\, r -
a^2}{r^2 + a^2} dt^2 +\fft{4ac\,r}{r^2+a^2} dt\,dz -
\fft{r^2+2a s\, r-a^2}{r^2+a^2} dz^2\Big)\nn\\
&& + (H_e H_m)^{\ft12}\Big(f [dr^2 + (r^2+a^2) (d\theta^2 + \sin^2\theta\,
d\phi^2)]
+f^{-1} a^2 (d\psi + n \cos\theta d\phi)^2\Big)\,,\nn\\
\varphi&=& \ft{1}{\sqrt2} \log \left(\fft{H_e}{H_m}\right)\,,\quad
F_\3=dA_\2\,,\quad A_\2 = \fft{1}{H_e} dt\wedge dz + 
q_m a\,\cos\theta d\phi\wedge d\psi\,,\nn\\
H_e&=& \alpha_e -\fft{q_e}{a} \arctan(\fft{r}{a})\,,\quad
H_m=\alpha_m - \fft{q_m}{a}\arctan(\fft{r}{a})\,,\quad
f=\alpha - n \arctan(\fft{r}{a})\,.
\eea
%%%%
In this solution, the level surface of the four-dimensional
tranverse is not spherical symmetric, but a squashed $S^3$.
This solution is effectively the same as the one discussed earlier,
and we shall not analyse it further.

         There is another dyonic string wormhole solution that
is spherically symmetric on the $S^3$, given by
%%%%
\bea
ds^2_6 &=&\fft{(H_e H_m)^{\ft12}}{\cos u}\Big [a^2 d\Omega_3^2 +
\fft{a^2 du^2}{4 \cos^2 u}\Big]\nn\\
&&+(H_e H_m)^{-\ft12} \Big(\cos v\, ( -dt^2 + dz^2) + 2 \sin v\,
dt dz\Big)\,,\nn\\
\varphi&=& \ft{1}{\sqrt2} \log \left(\fft{H_e}{H_m}\right)\,,\quad
F_\3= dt\wedge dz\wedge dH_e^{-1} + {*_6 (dt\wedge dz\wedge dH_m^{-1})}
\,,\nn\\
H_e&=& c_e -\fft{q_e}{a^2} u\,,\qquad
H_m=c_m - \fft{q_m}{a^2} u\,,\qquad
v=\ft12\sqrt3 (\pi -2u)\,.
\eea
%%%%%
The coordinate $u$ is related to the original $r$ coordinate by
%%%%
\be
\sin u = \fft{r\sqrt{r^2 + 2a^2}}{r^2 + a^2}\,.
\ee
%%%%
In order for the metric to be asymptotic to AdS$_3$ for $r\rightarrow
\infty$, it is necessary to choose $c_e=\pi q_e/(2a^2)$ and
$c_m=\pi q_m/(2a^2)$.  For simplicity, let us consider the
case with $q_e=q=q_m$.  We can reduce the solution on the three
sphere, with the reduction ansatz given by
%%%%
\be
ds^2_6=e^{2\alpha\varphi} ds_3^2 + e^{-\ft23\alpha\varphi} q\,d\Omega_\3^2\,,
\qquad F_3=2q (\Omega_\3 + q^{3/2}\, e^{4\alpha\varphi} \epsilon_\3)\,.
\ee
%%%% 
The resulting three-dimensional Langrangian is given by
%%%%
\be
{\cal L}_3 =\sqrt{-g} (R - \ft12(\del\varphi)^2 - V)\,,\qquad
V=-q^{-1} (6e^{\ft83\alpha\varphi} - 4 e^{4\alpha\varphi})\,.
\ee
%%%%
The corresponding $D=3$ metric is given by
%%%%%
\be
ds_3^2 = \fft{a^6}{\ell^6} \Big[\fft{H^4 a^2 du^2}{4\cos^6u} +
\fft{H^2}{\cos^3u} (\cos v (-dt^2 + dz^2) + 2 \sin v\, dt dz)\Big]\,.
\ee
%%%%
We now take the limit of $r\rightarrow \infty$, and compare the
solution with the metric in the Poincar\'e patch.  We find that the metric
becomes
%%%%
\be
ds_3^2 = \fft{\ell^2}{\rho^2} d\rho^2  + \Big(1 - \fft{3a^2}{4\rho^4}
\Big) \Big(-\fft{\rho^2}{\ell^2} dt^2 + \fft{\rho^2}{\ell^2} dz^2\Big) -
\fft{2\sqrt3 a^2}{\ell^2}\,dt\,dz + \cdots\,,
\ee
%%%%
where $\ell^2=q$.

Note that for general electric and magnetic charges $(q_e,q_m)$, we
have $\ell=\sqrt{q_eq_m}$.  Up to the order of $1/\rho^2$ in the cross-term and $1/\rho^4$ in the rest, 
this is precisely the metric given in (\ref{poin}).

%  The reason for $M=0$ in our solution is
%because we have set the boost parameter to zero.  The mass can be
%easily restored by performing a Lorentz boost in the $(t,z)$ plane.

\section{Conclusions}

   In this paper, we have constructed various examples of smooth 
Lorentzian-signature wormholes in supergravity theories.  In general,
the solutions are supported by gravity, a dilatonic scalar, and a 
$p$-form field strength, and 
the resulting 
wormhole connects two asymptotic regions that are locally flat.
Of particular
interest are the cases where only the metric and the $p$-form are involved.
In these non-dilatonic cases, the parameters in the solution can be adjusted
so that one of the asymptotic regions approaches AdS.

\section*{Acknowledgements}

   We are grateful to Gary Gibbons and Don Page for helpful discussions.
While this work was being completed, A.B. enjoyed the hospitality of the
Simons Workshop on Mathematics and Physics, and the KITP Miniprogram  
on Gauge Theory
and Langlands Duality, and thanks them for their support. 
C.N.P. is grateful to the Centre for Theoretical Cosmology, DAMTP,
Cambridge, for hospitality during the course of this work. The research  
of A.B. was supported
in part by DARPA and AFOSR through the grant FA9550-07-1-0543, by the  
National Science
Foundation under Grants No. PHY05-51164 and PHY-0505757, and by Texas  
A\&M University.  The research of C.N.P. is supported in part by 
DOE grant DE-FG03-95ER40917, and the Cambridge-Mitchell Collaboration. 

\appendix
\section{Conformal Mass for Asymptotically AdS Geometries}
\label{mass}

    In any spacetime that approaches AdS sufficiently rapidly at infinity, 
we can define conserved charges associated with each of the asymptotic
Killing vectors.  In particular, by taking the appropriate
asymptotically timelike Killing vector, we can calculate the total mass, or
energy, of the spacetime.  There are various ways in which this can be 
done (see, for example, \cite{chenlupope} for a discussion), but the 
simplest and
most straightforward is based on a procedure involving a calculation on
the conformal boundary of the spacetime, developed by 
Ashtekar, Magnon and Das \cite{ashmag,ashdas}.  
In this appendix, we show how this
AMD approach may be used to calculate the energy and the momentum of
the AdS wormhole solutions that we have constructed in this paper. 

     There are two types of boundary that are of particular interest when 
one considers an asymptotically AdS spacetime.  One of these is
the boundary at large radius in a global coordinate system, for which
the boundary topology is $\R\times$Sphere.  (It is assumed here that
we are working in the universal covering space CAdS of AdS, in which time 
ranges over the entire real line rather than being periodically identified.)
The other boundary of interest is the one that arises when one 
considers the Poincar\'e patch of AdS, for which the boundary is
just Minkowski spacetime.  

    Descriptions of the bulk AdS metrics in these two cases can be given in
a unified form, by writing the $D$-dimensional metric, which satisfies
$R_{\mu\nu}= -(D-1)\ell^{-2}\, g_{\mu\nu}$, as
%%%%%
\be
ds_D^2 = - w^2 dt^2 + \fft{dr^2}{w^2} + r^2 d\omega_{D-2}^2\,,\label{adsmet}
\ee
%%%%
where
%%%%
\be
d\omega_{D-2}^2 = \fft{du^2}{1-k^2 u^2} + u^2 d\Omega_{D-3}^2\,,\qquad
w^2=k^2 + \fft{r^2}{\ell^2}\,.\label{smet}
\ee 
%%%%%
For any non-vanishing $k$, $d\omega_{D-2}^2$ is the metric on a round
sphere $S^{D-2}$ of radius $1/k$, and (\ref{adsmet}) is a metric on
AdS$_D$ in global coordinates.  The scale size $k$ can be absorbed by
means of coordinate rescalings so that any non-zero $k$ can be set
equal to 1 without loss of generality.  If $k=0$, on the other hand,
the metric (\ref{adsmet}) instead describes the Poincar\'e patch of
AdS.\footnote{One can also take $k^2$ to be negative, in which case
  the metric (\ref{adsmet}) describes de Sitter spacetime, and
  $d\omega_{D-2}^2$ is the metric on a hyperboloid of constant
  negative curvature.  Including the possibility of negative $k^2$,
  one can always, by means of coordinate scalings, set $k^2$ to be 0,
  1 or $-1$, depending on whether it is initially zero, positive or
  negative.}

   The use of the AMD method to calculate the mass of various 
higher-dimensional asymptotically AdS black holes was described in
\cite{gibperpop} and \cite{chenlupope}.  In all these examples, the black hole
metrics were asymptotic to global AdS.  However, in the AdS wormhole
solutions that we have constructed in this paper, the asymptotic form
approaches the Poincar\'e patch of AdS.  It is instructive, therefore, 
first to check how the AMD calculation of the mass works in a 
simple $k=0$ example.

    The conformal boundary of AdS in either case can be brought in from
infinity by rescaling the metric with a conformal factor
%%%%
\be
\Omega = \fft{\ell}{r}\,,
\ee
%%%%
to give $\bar g_{\mu\nu} = \Omega^2 g_{\mu\nu}$.  Let $\bar
C^\mu{}_{\nu\rho\sigma}$ to be the Weyl tensor of the metric $\bar
g_{\mu\nu}$ and $\bar n_\mu=\del_\mu \Omega$.  The conserved charge
$Q[K]$ associated to the asympotic Killing vector $K$ is then given by
%%%%
\bea
Q[K]&=&\fft{\ell}{8\pi (D-3)}\oint_\Sigma \bar {\cal E}^\mu{}_\nu
K^\nu\, d\bar \Sigma_\mu\,,\nn\\
\bar{\cal E}^\mu{}_\nu &=& \ell^2 \Omega^{3-D} \bar n^\rho
\bar n^\sigma \bar C^\mu{}_{\rho\nu\sigma}\,.
\eea
%%%%%
In order to define the energy, one takes $K=\del/\del t$, to give
%%%%
\be
E=\fft{\ell}{8\pi (D-3)} \oint_\Sigma \bar{\cal E}^t{}_t d\bar\Sigma_t\,.
\ee
%%%%%%
In the case of our AdS wormhole solutions, we can also obtain
the linear momentum along the $z$ direction, by taking
$K=\del/\del z$, giving
%%%%
\be
P=\fft{1}{8\pi (D-3)} \oint_\Sigma \bar{\cal E}^t{}_z d\bar\Sigma_t\,.
\ee
%%%%%%

   A simple example that illustrates the calculation of the AMD mass
for both the $k=1$ and $k=0$ cases is provided by charged non-rotating
black holes in five-dimensional 
minimal gauged supergravity.  The solution can be written as
%%%%%
\bea
ds_5^2 &=& f\, H^{-2} \, dt^2 + H\, \Big(f^{-1}\, dt^2 + r^2 ds_3^2\Big)\,,
\nn\\
A &=& \sqrt 3\, (1-H^{-1}) \sqrt{\fft{\mu+ k^2 q}{q}}\, dt\,,
\eea
%%%%%
where
%%%%%
\be
f= k^2 -\fft{\mu}{r^2} + g^2 r^2 H^3 \,,\qquad H= 1 +\fft{q}{r^2}\,,
\ee
%%%%%
and 
%%%%%
\be
ds_3^2 = \fft{du^2}{1- k^2 u^2} + u^2 d\Omega_2^2\,.
\ee
%%%%%
Here $g=1/\ell$.  Note that $ds_3^2$ is a metric on a 3-sphere of
radius $k^{-1}$.  The solution is valid for any $k$, including $k=0$.

   Calculating the thermodynamic quantities, we find
%%%%%
\bea
S &=& \fft{\pi^2\, (q+r_+^2)^{3/2} }{2 k^3}\,,\qquad 
   T= \fft{k^2 r_+^4 + g^2 (2r_+^2 -q)(q+ r_+^2)^2}{2\pi r_+^2 
           (q+r_+^2)^{3/2} }\,,\nn\\
Q&=& \fft{\pi \sqrt{3q}\, \sqrt{q+r_+^2}\, 
         \sqrt{k^2 r_+^2 + g^2 (q+r_+^2)^2}}{k^3 r_+}\,,\qquad
  \Phi=\fft{\sqrt{3q}\,\sqrt{k^2 r_+^2 + g^2 (q+r_+^2)^2}}{
      4 r_+\, \sqrt{q+r_+^2}}\,,\nn\\
E &=& \fft{3\pi\Big[ k^2 r_+^2 (2q+r_+^2) + g^2 (q+r_+^2)^3\Big]}{8k^3 r_+^2}
\,,
\eea
%%%%%
where $r_+$ is the largest root of $f(r)=0$. $E$ here is calculated using the
AMD procedure.
These quantities satisfy the first law of thermodynamics,
%%%%%
\be
dE = T dS + \Phi dQ\,,
\ee
%%%%%
for any arbitrary constant value for $k$, including $k=0$.

    All of $E$, $S$ and $Q$ have a factor $k^3$ in the denominator.  This
is associated with the fact that $ds_3^2$ describes a 3-sphere of radius
$k^{-1}$.  In the limit where $k\rightarrow0$, corresponding to the
black hole with flat horizon, we should multiply
$E$, $S$ and $Q$ by $k^3$ before taking the limit, and interpret the
rescaled quantities as the energy, entropy and charge per unit
3-volume.  (Or else, re-interpret the metric $ds_3^2$ as being defined on
$T^3$, and so take $\int\sqrt{g_3} d^3 x$ to be the volume of the $T^3$.)


\begin{thebibliography}{99}

\bm{mald} J.M. Maldacena,
 {\it The large $N$ limit of superconformal field theories and supergravity,}
  Adv.\ Theor.\ Math.\ Phys.\  {\bf 2}, 231 (1998), 
  [Int.\ J.\ Theor.\ Phys.\  {\bf 38}, 1113 (1999)],
 hep-th/9711200.
%%CITATION = IJTPB,38,1113;%%

\bm{wit}
  E. Witten,
{\it Anti-de Sitter space and holography,}
  Adv.\ Theor.\ Math.\ Phys.\  {\bf 2}, 253 (1998),
  hep-th/9802150.
%%CITATION = 00203,2,253;%%

\bm{gkp} S.S. Gubser, I.R. Klebanov and A.M. Polyakov,
  {\it Gauge theory correlators from non-critical string theory,}
  Phys.\ Lett.\ {\bf B428}, 105 (1998), hep-th/9802109.
%%CITATION = PHLTA,B428,105;%%

\bibitem{fgpw}
  D.Z. Freedman, S.S. Gubser, K. Pilch and N.P. Warner,
  {\it Renormalization group flows from holography supersymmetry and a
  c-theorem,}
  Adv.\ Theor.\ Math.\ Phys.\  {\bf 3}, 363 (1999),
  hep-th/9904017.
%%CITATION = 00203,3,363;%%

\bm{llm}
  H. Lin, O. Lunin and J.M. Maldacena,
  {\it Bubbling AdS space and 1/2 BPS geometries,}
  JHEP {\bf 0410}, 025 (2004), hep-th/0409174.
%%CITATION = JHEPA,0410,025;%%

\bm{clpbubble}
  Z.W. Chong, H. L\"u and C.N. Pope,
  {\it BPS geometries and AdS bubbles,}
  Phys.\ Lett.\  {\bf B614}, 96 (2005),
hep-th/0412221.
%%CITATION = PHLTA,B614,96;%%

\bm{llpv}
  J.T. Liu, H. L\"u, C.N. Pope and J.F. Vazquez-Poritz,
  {\it New supersymmetric solutions of $N=2, D=5$ gauged supergravity with
  hyperscalars,}
  JHEP {\bf 0710}, 093 (2007)
  [arXiv:0705.2234 [hep-th]].
  %%CITATION = JHEPA,0710,093;%%

\bm{cglp}
  M. Cveti\v c, S.S. Gubser, H. L\"u and C.N. Pope,
  {\it Symmetric potentials of gauged supergravities in diverse dimensions
and Coulomb branch of gauge theories,}
  Phys.\ Rev.\  {\bf D62}, 086003 (2000),
hep-th/9909121.
%%CITATION = PHRVA,D62,086003;%%

\bibitem{klt} P. Kraus, F. Larsen and S.P. Trivedi,
 {\it The Coulomb branch of gauge theory from rotating branes,}
  JHEP {\bf 9903}, 003 (1999),
hep-th/9811120.
%%CITATION = JHEPA,9903,003;%%

\bm{gsww}
  G.J. Galloway, K. Schleich, D. Witt and E. Woolgar,
 {\it The AdS/CFT correspondence conjecture and topological censorship,}
  Phys.\ Lett.\ {\bf B505}, 255 (2001),
 hep-th/9912119.
%%CITATION = PHLTA,B505,255;%%

\bm{mm} J.M. Maldacena and L. Maoz,
{\it Wormholes in AdS,} JHEP {\bf 0402}, 053 (2004),
hep-th/0401024.
%%CITATION = JHEPA,0402,053;%%

\bm{bcpvv} E. Bergshoeff, A. Collinucci, A. Ploegh, S. Vandoren and
T. Van Riet,
{\it Non-extremal D-instantons and the AdS/CFT correspondence,}
  JHEP {\bf 0601}, 061 (2006), hep-th/0510048.
%%CITATION = JHEPA,0601,061;%%

\bm{hop} N. Arkani-Hamed, J. Orgera and J. Polchinski,
{\it Euclidean wormholes in string theory,}
JHEP {\bf 0712}, 018 (2007), arXiv:0705.2768 [hep-th].
%%CITATION = JHEPA,0712,018;%%

\bm{bd} A. Bergman and J. Distler, {\it Wormholes in maximal supergravity,}
arXiv:0707.3168 [hep-th].
%%CITATION = ARXIV:0707.3168;%%

\bm{bcptv} E. Bergshoeff, W. Chemissany, A. Ploegh,
M. Trigiante and T. Van Riet, {\it
Generating geodesic flows and supergravity solutions,}
arXiv:0806.2310 [hep-th].
%%CITATION = ARXIV:0806.2310;%%

\bm{lumei} H. L\"u and Jianwei Mei,
{\it Ricci flat and charged wormholes in five dimensions,}
arXiv:0806.3111 [hep-th].
%%CITATION = ARXIV:0806.3111;%%

\bm{cd} A. Chodos and S. Detweiler,
{\it Spherically symmetric solutions in five-dimensional general relativity,}
Gen.\ Rel.\ Grav.\  {\bf 14}, 879 (1982).
  %%CITATION = GRGVA,14,879;%%

\bm{aagc} M. Azreg-Ainou and G. Clement, {\it The geodesics of the
Kaluza-Klein wormhole soliton,} Gen.\ Rel.\ Grav.\ {\bf 22}, 1119
(1990).

\bm{aagccf} M. Azreg-Ainou, G. Clement, C.P. Constantinidis and
J.C. Fabris, {\it Electrostatic solutions in Kaluza-Klein theory:
Geometry and stability}, Grav.\ Cosmol.\ {\bf 6} 207 (2000),
gr-qc/9911107.

\bibitem{lpsoliton}
  H. L\"u and C.N. Pope,
  {\it p-brane solitons in maximal supergravities,}
  Nucl.\ Phys.\ {\bf B465}, 127 (1996),
hep-th/9512012.
%%CITATION = NUPHA,B465,127;%%

\bibitem{stainless}
  H. L\"u, C.N. Pope, E. Sezgin and K.S. Stelle,
  {\it Stainless super p-branes,}
  Nucl.\ Phys.\ {\bf B456}, 669 (1995), hep-th/9508042.
%%CITATION = NUPHA,B456,669;%%

\bm{instpaper} M.S. Bremer, M.J. Duff, H. L\"u, C.N. Pope and K.S. Stelle,
{\it Instanton cosmology and domain walls from M-theory and string theory,}
  Nucl. Phys.  B {\bf 543}, 321 (1999),
hep-th/9807051.
%%CITATION = NUPHA,B543,321;%%

%\bibitem{btz}
%  M. Banados, C. Teitelboim and J. Zanelli,
%  {\it The black hole in three-dimensional space-time,}
%  Phys.\ Rev.\ Lett.\  {\bf 69}, 1849 (1992),
%hep-th/9204099.
%%%CITATION = PRLTA,69,1849;%%

%\bibitem{bhtz}
%  M. Banados, M. Henneaux, C. Teitelboim and J. Zanelli,
%  {\it Geometry of the (2+1) black hole,}
%  Phys.\ Rev.\ {\bf D48}, 1506 (1993), gr-qc/9302012.
%%%CITATION = PHRVA,D48,1506;%%

%\bibitem{strom} A. Strominger,
%  {\it Black hole entropy from near-horizon microstates,}
%  JHEP {\bf 9802}, 009 (1998), hep-th/9712251.
%%%CITATION = JHEPA,9802,009;%%

\bibitem{BH}
  J.~D.~Brown and M.~Henneaux,
  {\it Central Charges in the Canonical Realization of Asymptotic Symmetries: An
  Example from Three-Dimensional Gravity},
  Commun.\ Math.\ Phys.\  {\bf 104}, 207 (1986).
  %%CITATION = CMPHA,104,207;%%

\bibitem{BK}
  V.~Balasubramanian and P.~Kraus,
  {\it A stress tensor for anti-de Sitter gravity},
  Commun.\ Math.\ Phys.\  {\bf 208}, 413 (1999)
  [arXiv:hep-th/9902121].
  %%CITATION = CMPHA,208,413;%%

\bibitem{Henningson:1998gx}
  M.~Henningson and K.~Skenderis,
  {\it The holographic Weyl anomaly},
  JHEP {\bf 9807}, 023 (1998)
  [arXiv:hep-th/9806087].
  %%CITATION = JHEPA,9807,023;%%


%%\bibitem{timemachine}
%%  M. Cveti\v c, G.W. Gibbons, H. L\"u and C.N. Pope,
%%{\it Rotating black holes in gauged supergravities: Thermodynamics,
%%  supersymmetric limits, topological solitons and time machines,}
%%hep-th/0504080.
%%CITATION = HEP-TH/0504080;%%

\bm{gubser} S.S. Gubser,
{\it Curvature singularities: The good, the bad, and the naked,}
  Adv. Theor. Math. Phys. {\bf 4}, 679 (2000),
hep-th/0002160.
%%CITATION = 00203,4,679;%%

\bm{chenlupope} W. Chen, H. L\"u and C.N. Pope,
 {\it Mass of rotating black holes in gauged supergravities,}
  Phys.\ Rev.\  {\bf D73}, 104036 (2006),
hep-th/0510081.
%%CITATION = PHRVA,D73,104036;%%

\bm{ashmag} A. Ashtekar and A. Magnon, {\it Asymptotically anti-de 
Sitter space-times}, Class Quant Grav {\bf 1}, L39 (1984).


\bm{ashdas} A. Ashtekar and S. Das, {\it Asymptotically anti-de Sitter
space-times: Conserved quantities}, Class. Quant. Grav. {\bf 17}, L17
(2000), hep-th/9911230.

\bm{gibperpop} G.W. Gibbons, M.J. Perry and C.N. Pope,
  {\it The first law of thermodynamics for Kerr-anti-de Sitter
black holes,}
  Class.\ Quant.\ Grav.\  {\bf 22}, 1503 (2005),
hep-th/0408217.
%%CITATION = CQGRD,22,1503;%%



\end{thebibliography}
\end{document}